\documentclass[10pt,twocolumn,superscriptaddress,amssymb,amsmath,aps,pra]{revtex4-1}
\usepackage{graphicx}

\newcommand*{\Scale}[2][4]{\scalebox{#1}{$#2$}}

\begin{document}

\title{Commutation-relation-preserving ladder operators for propagating optical fields in nonuniform lossy media}
\date{\today}
\author{Mikko Partanen}
\affiliation{Engineered Nanosystems group, School of Science, Aalto University, P.O. Box 12200, 00076 Aalto, Finland}
\author{Teppo H\"ayrynen}
\affiliation{Engineered Nanosystems group, School of Science, Aalto University, P.O. Box 12200, 00076 Aalto, Finland}
\affiliation{DTU Fotonik, Department of Photonics Engineering, Technical University of
Denmark, \O rsteds Plads, Building 343, DK-2800 Kongens Lyngby, Denmark}
\author{Jukka Tulkki}
\affiliation{Engineered Nanosystems group, School of Science, Aalto University, P.O. Box 12200, 00076 Aalto, Finland}
\author{Jani Oksanen}
\affiliation{Engineered Nanosystems group, School of Science, Aalto University, P.O. Box 12200, 00076 Aalto, Finland}

\begin{abstract}
We have recently developed a quantized fluctuational electrodynamics (QFED) formalism
to describe the quantum aspects of local thermal balance formation and to 
formulate the electromagnetic field ladder operators so that they no longer exhibit
the anomalies reported for resonant structures. Here we show how the QFED can be used
to resolve between the left and right propagating fields to bridge the QFED
and the quantum optical input-output relations commonly used to describe selected
quantum aspects of resonators. The generalized model introduces a density
of states concept describing
interference effects, which is instrumental in allowing an
unambiguous separation of the fields and related quantum operators
into left and right propagating parts. In addition to providing
insight on the quantum treatment of interference, our results also
provide the conclusive resolution of the long-standing enigma of the
anomalous commutation relations of partially confined propagating fields.
\end{abstract}

\maketitle

\section{Introduction}

It has recently been suggested
\cite{Gauvin2014,Collette2013,Partanen2014a,Partanen2014c}
that in contrast to earlier predictions
the commutation relations of photon ladder operators
have directly measurable physical significance.
This is especially interesting in the case of
resonant structures
where conventional theoretical descriptions have been shown to lead to
anomalous commutation relations
of the general form
$[\hat a,\hat a^\dag]=\Lambda\neq1$,
instead of the canonical form with $\Lambda=1$
\cite{Ueda1994,Raymer2013,Barnett1996,Aiello2000,Stefano2000}.
The anomaly is a
direct consequence of the conventional piecewise normalization of the optical
modes in respective homogeneous regions and of
the interference effects coupling the modes propagating
in different directions.
Originally these anomalies were argued to bear no physical significance,
but recently it has been shown that the anomalous commutation relations
lead to the existence of a threshold for second harmonic generation
when it occurs inside microcavities \cite{Gauvin2014,Collette2013}.
In addition, anomalies in the commutation relations
have also been shown to prevent systematic description of the local
thermal balance between the field and interacting media \cite{Partanen2014a,Partanen2014c,Partanen2014b}.
Experimental measurements of the onset of the second harmonic
generation or the thermal balance formation in microcavities could therefore
confirm the theoretical predictions that the conventional mode normalization
introducing these anomalies is not sufficient for field quantization of resonant structures.
Here, we develop a field quantization approach that (1) fully eliminates
the anomalies for propagating fields, (2) bridges the classical propagating wave and
commutation-relation-preserving quantum descriptions, and (3) allows formulating
conceptually simple models for optical energy transfer and the
formation of thermal balance in interfering nanostructures.

One of the most widely used quantization approaches for describing spatial field evolution
in resonant structures is the input-output relation formalism of the photon creation
and annihilation operators. The formalism was originally developed for dispersionless
and lossless media \cite{Knoll1987} and later extended for lossy and dispersive
dielectrics by several groups
\cite{Knoll1991,Allen1992,Huttner1992,Barnett1995,Matloob1995,Matloob1996}.
The quantization procedures studied, e.g., by Barnett \emph{et al.} \cite{Barnett1995}
clearly highlight that the noise and field operators in nonuniform systems
are position dependent and that the vector potential and electric-field operators
obey the well-known canonical commutation relation
as expected \cite{Barnett1995,Matloob1995}. The canonical commutation relations
in these early models did not, however, extend to the ladder operators which were
found to exhibit anomalies in resonant structures \cite{Ueda1994}.
The anomalous commutation relations of the ladder operators were later studied in several works
\cite{Raymer2013,Barnett1996,Aiello2000,Stefano2000} but no clear resolution for the
anomalies was found. Instead it was concluded that the anomalies
as well as the exact form of the ladder operators within resonant structures were
irrelevant as long as the field commutation relations and classical field
quantities were well defined. However, this made it impossible to fully quantize the optical
fields in resonant structures.

To shed more light on the anomalous commutation relations,
we have very recently developed a quantized fluctuational electrodynamics (QFED) scheme based on
generalizing the fluctuational electrodynamics to quantum optical fields
\cite{Partanen2014c,Partanen2014a,Partanen2014b}. 
Using the QFED approach we were able to formulate the 
canonical commutation relations preserving ladder and photon-number operators
for the total electromagnetic (EM) field
\cite{Partanen2014a,Partanen2014b,Partanen2014c}.
However, even in the QFED framework, it has not been evident how to separate
the ladder and photon-number operators to left and right propagating parts,
which is also essential for the final resolution of the anomalies and
for bridging the classical propagating wave descriptions and the
commutation-relation-preserving quantum descriptions.
In this work, we show that the QFED can be extended
to resolve between the left and right
propagating fields,
fully preserving the canonical commutations with $\Lambda=1$
also for the left and right propagating field
ladder operators. As it turns out, the separation
to the left and right propagating fields becomes possible
and conceptually simple when
one introduces a new density of states concept describing the fundamentally
important interference effects.
The added insight obtained using the introduced concepts provides a more detailed
understanding of the quantization of optical fields in complex surroundings, and
can be used, e.g., for finding simple photon-number-based expressions for the
quantum optical Poynting vector as well as more detailed description
of quantized energy flow in resonant structures.

\section{\label{sec:theory}Field quantization}

\subsection{\label{sec:operators}Photon operators}

In contrast to previous approaches,
the requirement of the QFED is the preservation of the local
canonical commutation relation
$[\hat a(x,\omega),\hat a^\dag(x,\omega')]=\delta(\omega-\omega')$
also in resonant media \cite{Partanen2014a,Partanen2014b}. This requirement leads to conceptually simple definitions for the
position-dependent ladder and photon-number operators
as a weighted sum over the incident fields and the noise.
For the expectation value of
the photon-number operator, for instance, this weighted sum reads as \cite{Partanen2014c}
\begin{equation}
 \langle\hat n(x,\omega)\rangle =\frac{\int_{-\infty}^\infty\rho_\mathrm{NL}(x,\omega,x')\langle\hat\eta(x',\omega)\rangle dx'}{\int_{-\infty}^\infty\rho_\mathrm{NL}(x,\omega,x')dx'},
 \label{eq:totaln}
\end{equation}
where $\langle\hat \eta(x,\omega)\rangle$ is the source field
photon-number expectation value which for thermal fields is
given by the Bose-Einstein distribution
$\langle\hat \eta(x',\omega)\rangle=1/[e^{\hbar\omega/(k_\mathrm{B}T(x'))}-1]$ with the
position-dependent temperature of the medium given by $T(x')$.
The weighting coefficient $\rho_\mathrm{NL}(x,\omega,x')$
in Eq.~\eqref{eq:totaln} is given by
\begin{align}
 \rho_\mathrm{NL}(x,\omega,x') & =\frac{\omega^3|\varepsilon(x,\omega)|}{\pi c^4S}\,\varepsilon_\mathrm{i}(x',\omega)\nonumber\\
 &\hspace{0.5cm}\times\Big(|G(x,\omega,x')|^2+\Big|\frac{\partial G(x,\omega,x')}{k(x,\omega)\partial x}\Big|^2\Big),
 \label{eq:nldos}
\end{align}
where $c$ is the speed of light in vacuum,
$S$ is the area of quantization in the $y$-$z$ plane,
$\varepsilon(x,\omega)=n(x,\omega)^2$ is the relative electric permittivity of a
nonmagnetic medium with refractive index
$n(x,\omega)$ and $\varepsilon_\mathrm{i}(x,\omega)$ is its imaginary part,
$k(x,\omega)=\omega n(x,\omega)/c$ is the wave number,
and $G(x,\omega,x')$ is the Green's function of the Helmholtz equation
given for selected layered structures in Ref.~\citenum{Partanen2014a}.
The quantity $\rho_\mathrm{NL}(x,\omega,x')$ is here
referred to as the nonlocal density of states (NLDOS)
since it highlights the nonlocal origin of the local density of states (LDOS)
$\rho(x,\omega)$ given as
\begin{equation}
 \rho(x,\omega) =\int_{-\infty}^\infty\rho_\mathrm{NL}(x,\omega,x')dx',
 \label{eq:uldos}
\end{equation}
and appearing in the denominator of Eq.~\eqref{eq:totaln}.
After integration, the LDOS can also be expressed in the more familiar form
in terms of the imaginary part of the Green's function \cite{Partanen2014c}.
Using this definition, the NLDOS $\rho_\mathrm{NL}(x,\omega,x')$
accounts for both the electric-field (term $|G|^2$)
and the magnetic-field (term $|\partial G/(k\partial x)|^2$) contributions.

\subsection{Quantum optical Poynting vector}

The quantum optical Poynting vector $\hat{S}(x,t)$ that will be used as a starting point for separating the
field components is
defined in terms of the positive ($+$) and negative ($-$) frequency parts of the electric-
and magnetic-field operators $\hat{E}(x,t)$ and $\hat{B}(x,t)$
as $\hat{S}(x,t)=\varepsilon_0c^2[\hat{E}^-(x,t)\hat{B}^+(x,t)+\hat{B}^-(x,t)\hat{E}^+(x,t)]$
\cite{Janowicz2003,Loudon2000}.
Using the QFED framework and substituting the electric- and
magnetic-field operators as given in Ref.~\citenum{Partanen2014a},
we write the Poynting vector expectation value at angular frequency $\omega$ as
\begin{align}
 \langle\hat S(x,t)\rangle_\omega & =\hbar\omega v(x,\omega)\int_{-\infty}^\infty\rho_\mathrm{IF}(x,\omega,x')\langle\hat\eta(x',\omega)\rangle dx'
 \label{eq:poynting1},
\end{align}
where $v(x,\omega)=c/n_\mathrm{r}(x,\omega)$ is the energy propagation velocity,
$n_\mathrm{r}(x,\omega)$ is the real part of the refractive index $n(x,\omega)=\sqrt{\varepsilon(x,\omega)}$, and
\begin{align}
 \rho_\mathrm{IF}(x,\omega,x') & =\frac{2\omega n_\mathrm{r}(x,\omega)}{\pi c^3S}\,\varepsilon_\mathrm{i}(x',\omega)\nonumber\\
 &\hspace{0.5cm}\times\mathrm{Re}\Big(i\omega G(x,\omega,x')\frac{\partial G^*(x,\omega,x')}{\partial x}\Big).
 \label{eq:interfdens}
\end{align}
The quantity $\rho_\mathrm{IF}(x,\omega,x')$ at field point $x$
essentially describes the contributions of the left and right propagating
fields originating from the source point $x'$. The term $\rho_\mathrm{IF}(x,\omega,x')$
fully accounts for the reflections,
losses, and interference and is closely related to the concepts of LDOS and NLDOS.
Therefore, we refer to it as the interference density of states (IFDOS). In contrast to the NLDOS,
the integral of the IFDOS with respect to $x'$ is always zero as required, e.g., by the fact
that in a medium in equilibrium, there is no net power flow, i.e.,
$\langle\hat S(x,t)\rangle_\omega=0$ when $\langle\hat\eta(x',\omega)\rangle$ is constant.

\subsection{Left and right propagating fields}

To generalize the LDOS and photon-number concepts in Refs.~\citenum{Partanen2014a} and \citenum{Partanen2014c}
and Eq.~\eqref{eq:totaln} to separately account for the left and right propagating fields, we write
the left and right propagating field Poynting vector expectation values
$\langle\hat S_+(x,\omega)\rangle$ and $\langle\hat S_-(x,\omega)\rangle$ as
$\langle\hat S_\pm(x,\omega)\rangle_\omega=\hbar\omega v(x,\omega)\rho_\pm(x,\omega)\big(\langle\hat n_\pm(x,\omega)\rangle+\frac{1}{2}\big)$,
where $\rho_\pm(x,\omega)$ and $\langle\hat n_\pm(x,\omega)\rangle$ are the left and right
propagating field LDOSs and photon numbers to be determined, and the
term one half describes the zero-point fluctuation current.
The left and right propagating photon numbers must additionally
satisfy two equations: the total Poynting vector must be given by
\begin{align}
 \langle\hat S(x,t)\rangle_\omega &
=\hbar\omega v(x,\omega)\rho_+(x,\omega)\Big(\langle\hat n_+(x,\omega)\rangle+\frac{1}{2}\Big)\nonumber\\
&\hspace{0.5cm}-\hbar\omega v(x,\omega)\rho_-(x,\omega)\Big(\langle\hat n_-(x,\omega)\rangle+\frac{1}{2}\Big)
 \label{eq:poynting2},
\end{align}
and the total energy density $\langle\hat u(x,t)\rangle_\omega=\hbar\omega\rho(x,\omega)\big(\langle\hat n(x,\omega)\rangle+\frac{1}{2}\big)$
\cite{Partanen2014c} by
\begin{align}
 \langle\hat u(x,t)\rangle_\omega &= \hbar\omega\rho_+(x,\omega)\Big(\langle\hat n_+(x,\omega)\rangle+\frac{1}{2}\Big)\nonumber\\
 &\hspace{0.5cm}+\hbar\omega\rho_-(x,\omega)\Big(\langle\hat n_-(x,\omega)\rangle+\frac{1}{2}\Big).
 \label{eq:edensity1}
\end{align}
At zero temperature, where $\langle\hat n_+(x,\omega)\rangle=\langle\hat n_-(x,\omega)\rangle=0$,
the Poynting vector is zero and thus $\rho_+(x,\omega)=\rho_-(x,\omega)$ in Eq.~\eqref{eq:poynting2}.
Respectively, Eq.~\eqref{eq:edensity1}
at zero temperature leads to the relation $\rho_+(x,\omega)+\rho_-(x,\omega)=\rho(x,\omega)$.
Together, these conditions uniquely define the left and
right propagating LDOSs in terms of the total LDOS as
$\rho_+(x,\omega)=\rho_-(x,\omega)=\rho(x,\omega)/2$.

Using the above local density of states relations, we can uniquely solve the left and right propagating
photon numbers from Eqs.~\eqref{eq:poynting2} and \eqref{eq:edensity1} as
$\langle\hat n_\pm(x,\omega)\rangle=[\hbar\omega\rho(x,\omega)]^{-1}[\langle\hat u(x,t)\rangle_\omega\pm\langle\hat S(x,t)\rangle_\omega/v(x,\omega)]-1/2$.
In terms of the source field photon number this corresponds to
% \begin{equation}
%  \langle\hat n_\pm(x,\omega)\rangle =\frac{\int_{-\infty}^\infty[\rho_\mathrm{NL}(x,\omega,x')\pm\rho_\mathrm{IF}(x,\omega,x')]\langle\hat\eta(x',\omega)\rangle dx'}{\int_{-\infty}^\infty[\rho_\mathrm{NL}(x,\omega,x')\pm\rho_\mathrm{IF}(x,\omega,x')]dx'}.
%  \label{eq:directionaln}
% \end{equation}
\begin{equation}
 \Scale[0.95]{\displaystyle\langle\hat n_\pm(x,\omega)\rangle =\frac{\int_{-\infty}^\infty[\rho_\mathrm{NL}(x,\omega,x')\pm\rho_\mathrm{IF}(x,\omega,x')]\langle\hat\eta(x',\omega)\rangle dx'}{\int_{-\infty}^\infty[\rho_\mathrm{NL}(x,\omega,x')\pm\rho_\mathrm{IF}(x,\omega,x')]dx'}.}
 \label{eq:directionaln}
\end{equation}
Equation \eqref{eq:directionaln} shows that the propagating field photon-number expectation values
are also obtained as a weighted sum of the source field values, but the weight factor now
includes an additional term describing the interference and propagation direction.
In the denominator, one can neglect $\rho_\mathrm{IF}(x,\omega,x')$ as it integrates to zero
indicating that the denominator is simply equal to the LDOS in Eq.~\eqref{eq:uldos}.

Above we have only focused on the photon-number expectation values that can be directly
extracted from the Poynting vector. To find the corresponding ladder and
photon-number operators in the QFED we will further investigate the forms of
the photon annihilation operators $\hat a_+(x,\omega)$ and $\hat a_-(x,\omega)$
that lead to the expectation values in Eq.~\eqref{eq:directionaln} and
fulfill the canonical commutation relations.
The photon annihilation operators fulfilling these conditions are of the form
% \begin{equation}
%  \hat a_\pm(x,\omega)=\frac{1}{\sqrt{\rho(x,\omega)}}\int_{-\infty}^\infty\!e^{i(\phi\pm\pi/4)}\sqrt{\rho_\mathrm{NL}(x,\omega,x')\pm\rho_\mathrm{IF}(x,\omega,x')}\,\hat f(x',\omega)dx',
%  \label{eq:directionala}
% \end{equation}
\begin{align}
 \hat a_\pm(x,\omega) &=\frac{1}{\sqrt{\rho(x,\omega)}}\int_{-\infty}^\infty\!e^{i(\phi\pm\pi/4)}\nonumber\\
 &\hspace{0.5cm}\times\sqrt{\rho_\mathrm{NL}(x,\omega,x')\pm\rho_\mathrm{IF}(x,\omega,x')}\,\hat f(x',\omega)dx',
 \label{eq:directionala}
\end{align}
where $\hat f(x',\omega)$ is a bosonic source field operator obeying the canonical commutation
relation $[\hat f(x,\omega),\hat f^\dag(x',\omega')] = \delta(x-x')\delta(\omega-\omega')$
and which is related to the source field photon number as
$\langle\hat\eta(x',\omega)\rangle=\int\langle\hat f^\dag(x',\omega)\hat f(x'',\omega')\rangle dx''d\omega'$
\cite{Partanen2014a,Partanen2014c,Partanen2014b}.
The phase factor $\phi$ is in principle arbitrary and it does not play
a role in our calculations as it cancels in the commutators.
The total field annihilation operator $\hat a(x,\omega)$
is given by the sum
$\hat a(x,\omega)=\frac{1}{\sqrt{2}}[\hat a_+(x,\omega)+\hat a_-(x,\omega)]$.
It is straightforward to check that the left and right propagating field
annihilation operators in Eq.~\eqref{eq:directionala} also obey the commutation relation
of the form
$[\hat a_\pm(x,\omega),\hat a_\pm^\dag(x,\omega')] = \delta(\omega-\omega')$.
With these choices, however, the cross-commutators become nonzero as
$[\hat a_\pm(x,\omega),\hat a_\mp^\dag(x,\omega')]\neq0$
due to the coupling of the left and right propagating fields
originating from the same source points by the reflecting interfaces.
This cross-commutator form is intuitively reasonable and does
not appear to present any complications as the only  commutation relations directly
linked to the studied physical observables are the self-commutators.

As shown by Eqs.~\eqref{eq:directionaln} and \eqref{eq:directionala} it is necessary to
separately account for all the individual source points and their mutual interference
to arrive to the correctly commutating operator forms.
Similar book keeping is also present in classical fluctuational electrodynamics (FED).
In contrast to the FED, however, to describe the quantum features
the photon ladder and number operators need to be
renormalized to fully satisfy the commutation relations.

In addition to describing the total energy density and energy flow
presented in Eqs.~\eqref{eq:poynting1}, \eqref{eq:poynting2}, and \eqref{eq:edensity1},
the QFED formalism is also capable of separating the total field photon
numbers to their local electric- and magnetic-field equivalents
$\langle\hat n_\mathrm{e}(x,\omega)\rangle$ and
$\langle\hat n_\mathrm{m}(x,\omega)\rangle$ that are responsible
for direct interactions with materials
and determine, e.g., the self-consistent local temperature of the interacting
media as discussed in Refs.~\citenum{Partanen2014c} and \citenum{Partanen2014a}. Essentially these
electric- or magnetic-field specific quantities and the corresponding LDOSs
can be obtained by using Eqs.~\eqref{eq:totaln} and \eqref{eq:uldos}
when only the electric- or magnetic-field term in the NLDOS
in Eq.~\eqref{eq:nldos} is taken into account \cite{Partanen2014c}.
The electric- and magnetic-field specific quantities were
previously shown
to have quite distinct properties as compared to the total field quantities and to
include, e.g., oscillations in the field temperatures \cite{Partanen2014c,Partanen2014a}.
In the propagating operator formalism, however, the electric- and magnetic-field specific
ladder operators are again united with the total propagating field operators
because the direct interference effects between the left and
right propagating fields have been eliminated when projecting the
ladder operators to the left and right propagating operators.
This fully agrees with our previous results \cite{Partanen2014c,Partanen2014a}:
in the present formalism the separation to left and right propagating
fields also fully separates the interference effects from the local
fields, whereas the formalism simultaneously capturing both left and
right propagating fields in a single term must also capture the interference
effects.

\section{Results}

\begin{figure}
\includegraphics[width=0.48\textwidth]{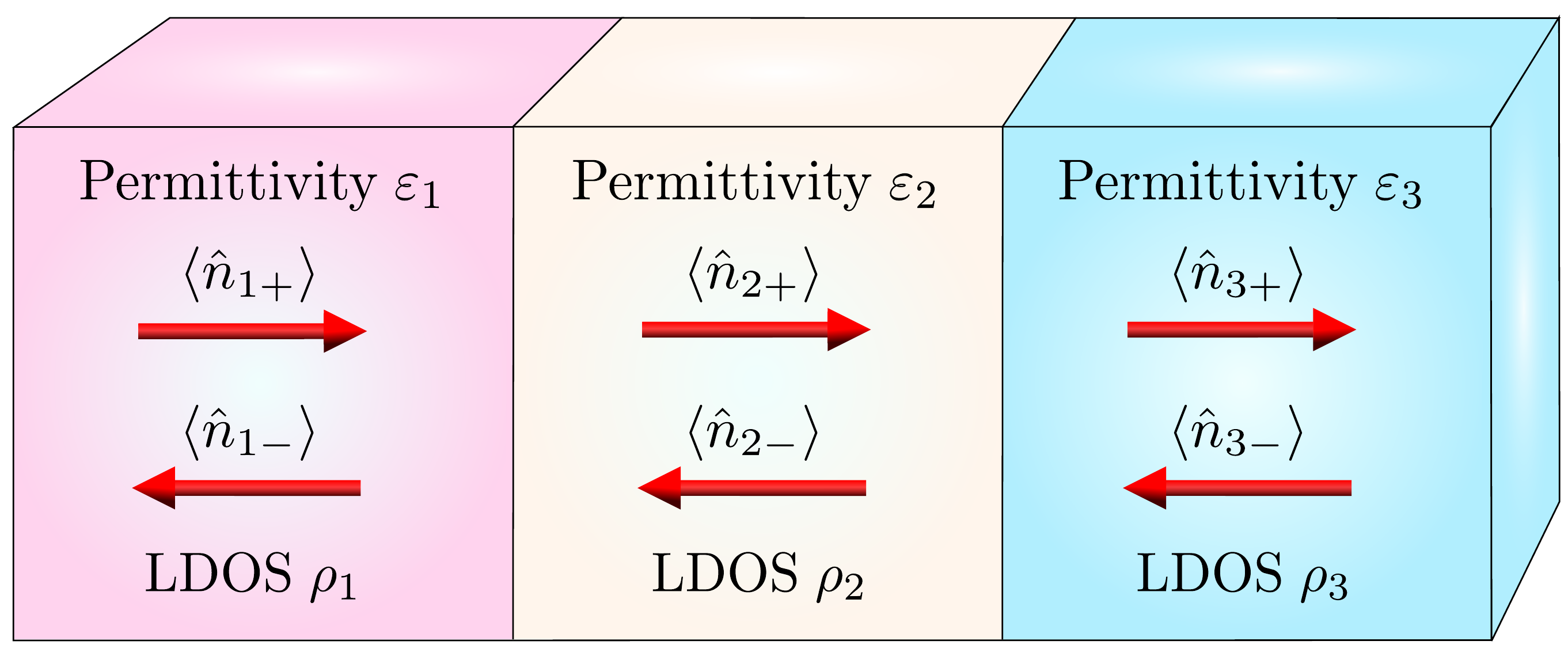}
\caption{\label{fig:cavity}(Color online) Optical cavity consisting of three homogeneous layers.
We calculate the left and right propagating field photon-number
expectation values in each layer.}
\end{figure}

To better illustrate the physical implications of the presented concepts
we briefly discuss the properties of
photon numbers of the left and right propagating fields and compare them
to the corresponding total field photon number
in an optical cavity consisting of three homogeneous layers as
illustrated in Fig.~\ref{fig:cavity}.

\subsection{Lossless cavity structure}

In a lossless configuration, the left and right propagating field photon numbers are piecewise
continuous and only depend on the cavity geometry and the input fields $\langle\hat n_{1+}\rangle$
and $\langle\hat n_{3-}\rangle$ incident from the left and right.
In different regions, they can be written as
\begin{align}
 \langle\hat n_{1-}\rangle & =\textstyle|\mathcal{R}_1|^2\langle\hat n_{1+}\rangle+\sqrt{\varepsilon_1/\varepsilon_3}\,|\mathcal{T}_1'\mathcal{T}_2'|^2\langle\hat n_{3-}\rangle,\nonumber\\
 \langle\hat n_{2+}\rangle & =\frac{\sqrt{\varepsilon_2/\varepsilon_1}\,|\mathcal{T}_1|^2\langle\hat n_{1+}\rangle+\sqrt{\varepsilon_2/\varepsilon_3}\,|\mathcal{T}_2'\mathcal{R}_1'|^2\langle\hat n_{3-}\rangle}{\mathrm{Re}[1+2\mathcal{R}_1'\mathcal{R}_2\nu_2e^{2ik_2d_2}]},\nonumber\\
 \langle\hat n_{2-}\rangle & =\frac{\sqrt{\varepsilon_2/\varepsilon_1}\,|\mathcal{T}_1\mathcal{R}_2|^2\langle\hat n_{1+}\rangle+\sqrt{\varepsilon_2/\varepsilon_3}\,|\mathcal{T}_2'|^2\langle\hat n_{3-}\rangle}{\mathrm{Re}[1+2\mathcal{R}_1'\mathcal{R}_2\nu_2e^{2ik_2d_2}]},\nonumber\\
 \langle\hat n_{3+}\rangle & =\textstyle\sqrt{\varepsilon_3/\varepsilon_1}\,|\mathcal{T}_1\mathcal{T}_2|^2\langle\hat n_{1+}\rangle+\textstyle|\mathcal{R}_2'|^2\langle\hat n_{3-}\rangle,
 \label{eq:photons}
\end{align}
where $d_2$ is the cavity thickness, $k_2$ is the wave number inside the cavity,
$\nu_2=1/(1+r_1r_2e^{2ik_2d_2})$, $\mathcal{R}_1=(r_1+r_2e^{2ik_2d_2})\nu_2$,
$\mathcal{R}_2=r_2$,
$\mathcal{T}_1=t_1\nu_2$, $\mathcal{T}_2=t_2$, $\mathcal{R}_1'=r_1'$,
$\mathcal{R}_2'=(r_2'+r_1'e^{2ik_2d_2})\nu_2$, $\mathcal{T}_1'=t_1'$,
and $\mathcal{T}_2'=t_2'\nu_2$ with
the conventional single interface
Fresnel reflection and transmission coefficients for left incidence
$r_i$ and $t_i$, $i\in\{1,2\}$, and right incidence $r_i'$ and $t_i'$, $i\in\{1,2\}$.
In contrast, e.g., to the electric-field values where resonance effects can substantially
increase the field magnitude inside a resonator,
the photon-number values inside the cavity and at the outputs in Eq.~\eqref{eq:photons}
are always between the input field photon numbers. This also ensures that
in global thermal equilibrium all the photon numbers are equal and no photon-number
accumulation can occur inside the cavity.

\begin{figure}
\includegraphics[width=0.43\textwidth]{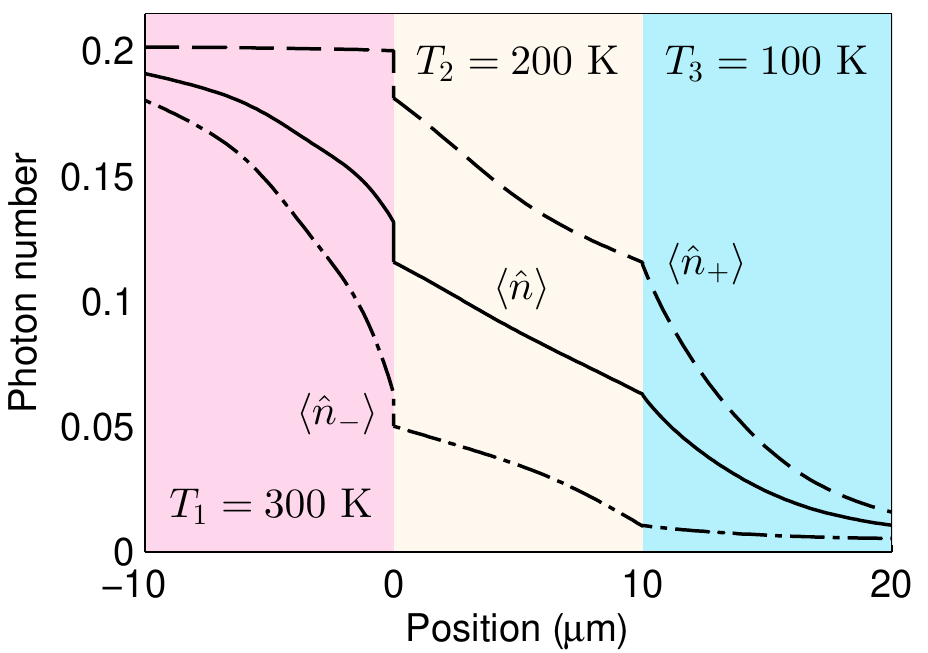}
\caption{\label{fig:photons}(Color online) Left and right propagating
photon numbers $\langle\hat n_-\rangle$ and $\langle\hat n_+\rangle$
and the total photon number $\langle\hat n\rangle$ in a lossy cavity structure
at the first resonant energy $\hbar\omega=0.046$ eV ($\lambda=26.9$ $\mu$m).
The media from left to right have refractive indices $\sqrt{\varepsilon_1}=2.5+0.4i$,
$\sqrt{\varepsilon_2}=1.2+0.2i$, and $\sqrt{\varepsilon_3}=1.5+0.5i$, and source field temperatures
$T_1=300$ K, $T_2=200$ K, and $T_3=100$ K.}
\end{figure}

\begin{figure*}
\includegraphics[width=0.8\textwidth]{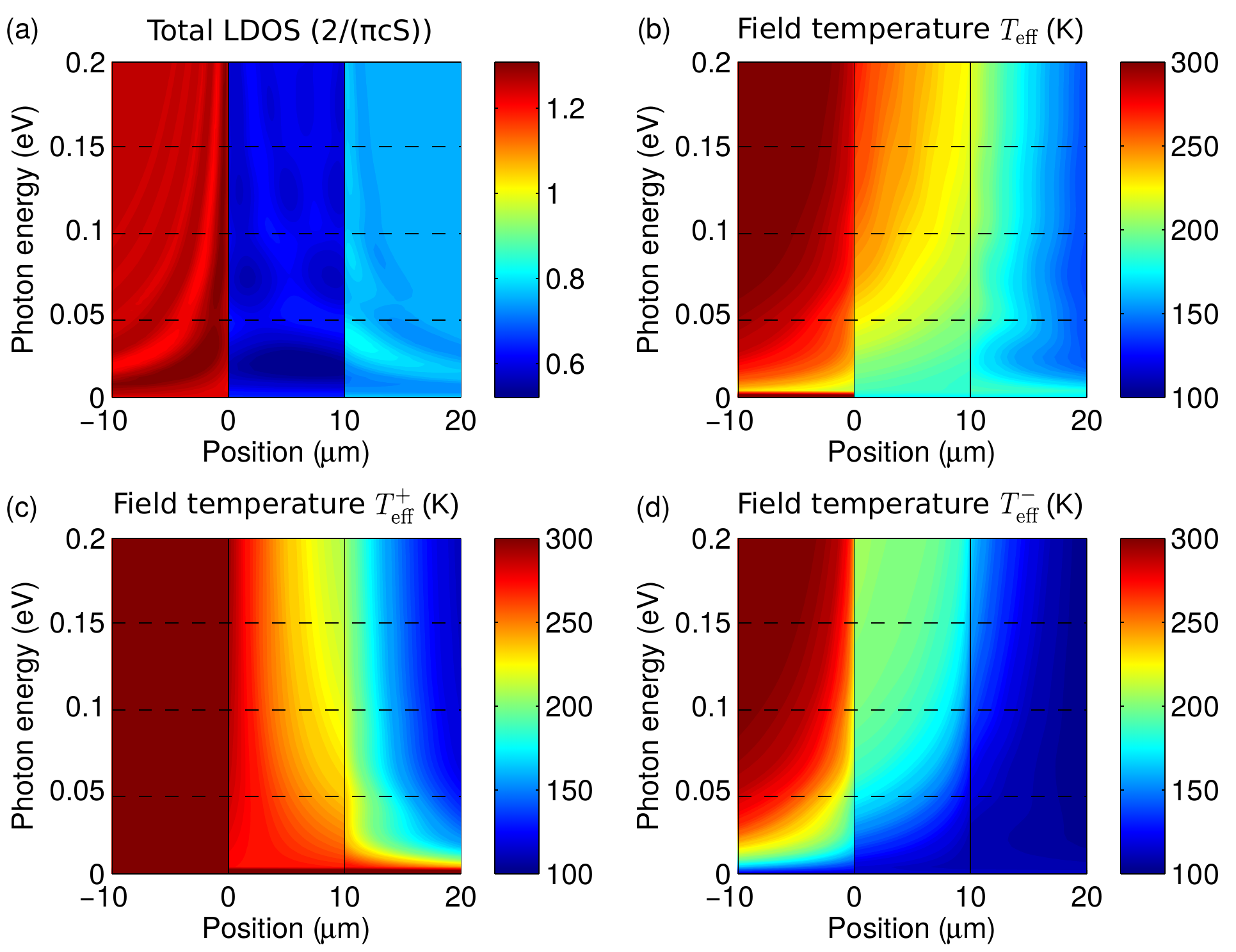}
\caption{\label{fig:temp}(Color online) (a) Total LDOS,
(b) $T_\mathrm{eff}$, (c)
$T_\mathrm{eff}^+$, and (d) $T_\mathrm{eff}^-$
in a lossy cavity structure.
The media from left to right have refractive indices $\sqrt{\varepsilon_1}=2.5+0.4i$,
$\sqrt{\varepsilon_2}=1.2+0.2i$, and $\sqrt{\varepsilon_3}=1.5+0.5i$, and layer temperatures
$T_1=300$ K, $T_2=200$ K, and $T_3=100$ K.
Solid lines denote the boundaries of the cavity and dashed lines denote resonant energies.
The LDOS is given in the units of $2/(\pi c S)$.}
\end{figure*}

\subsection{Lossy cavity structure}

In a lossy structure the photon numbers are no longer piecewise constant
and all material points can act as field sources through the
source field $\langle\hat\eta(x',\omega)\rangle$, which is related to material temperature.
To illustrate this, we study a lossy cavity structure, where the
refractive indices of the media are $\sqrt{\varepsilon_1}=2.5+0.4i$,
$\sqrt{\varepsilon_2}=1.2+0.2i$, and $\sqrt{\varepsilon_3}=1.5+0.5i$, and the layer
temperatures are $T_1=300$ K, $T_2=200$ K, and $T_3=100$ K.
Here the layer temperatures are set to constant predefined values
for simplicity even if the QFED formalism also allows calculating
the in-cavity temperature distribution self-consistently
if the studied layers were to be considered as thermal insulators
\cite{Partanen2014c,Partanen2014b}.
Figure \ref{fig:photons} shows the
the total, right propagating, and left propagating photon numbers
$\langle\hat n\rangle$, $\langle\hat n_+\rangle$, and $\langle\hat n_-\rangle$
as a function of position at the first cavity resonance
$\hbar\omega=0.046$ eV ($\lambda=26.9$ $\mu$m), where the
layer temperatures correspond to steady-state photon numbers
0.20, 0.074, and 0.0048.
The photon numbers are highest at the leftmost medium at
$T_1=300$ K and decrease towards the rightmost medium at
$T_3=100$ K. The right propagating photon number notably
decreases at and after the first interface due to reflection and thermalization,
eventually reaching equilibrium with the
lossy medium in the rightmost layer. The left propagating
photon number notably changes at the interfaces and
in the middle and leftmost layers.
It can be also clearly seen that the total photon number is
the average of the left and right
propagating photon numbers as expected, since the
photon number essentially describes the average photon
number in the collection of optical modes under study.

As the photon-number expectation value depends strongly on the frequency,
it is convenient to illustrate the results by using
the effective field temperature that is defined in terms of
the photon-number expectation value as
$T_\mathrm{eff}(x,\omega)=\hbar\omega/\big(k_\mathrm{B}\ln[1+1/\langle\hat n(x,\omega)\rangle]\big)$
\cite{Partanen2014c,Partanen2014b}.
This corresponds to the steady-state temperature of a small temperature probe
interacting only with a single mode \cite{Partanen2014c}.
Figure \ref{fig:temp} shows the
total LDOS and field temperatures
corresponding to the total, right propagating, and left propagating fields
as a function of position and photon energy.
In contrast to the case of a lossless structure, the field quantities
are position dependent.
The total LDOS in Fig.~\ref{fig:temp}(a) is also
oscillatory inside the cavity
and reaches its maxima at resonant
energies $\hbar\omega=0.046$ eV ($\lambda=26.9$ $\mu$m),
$\hbar\omega=0.097$ eV ($\lambda=12.7$ $\mu$m), and $\hbar\omega=0.150$ eV ($\lambda=8.29$ $\mu$m).
Also in the left- and rightmost layers, the total LDOS
is position dependent and oscillatory near interfaces.
The oscillations of the LDOS follow from
the interference effects combined with the material
polarizability in analogy with the Purcell effect \cite{Partanen2014c}.

Despite the oscillations in the LDOS, the total effective field temperature $T_\mathrm{eff}$
in Fig.~\ref{fig:temp}(b) and the effective field temperatures $T_\mathrm{eff}^+$ and $T_\mathrm{eff}^-$
corresponding to the right and left propagating fields in Figs.~\ref{fig:temp}(c)
and \ref{fig:temp}(d) decrease towards the right medium at lower temperature
similar to the photon numbers in Fig.~\ref{fig:photons}. On the left and right
$T_\mathrm{eff}$, $T_\mathrm{eff}^+$, and $T_\mathrm{eff}^-$ also asymptotically
approach equilibrium values corresponding to material temperatures.
When compared to $T_\mathrm{eff}^+$,
the magnitude of $T_\mathrm{eff}^-$ is
everywhere lower since the source field temperature on the right is lower than
the source field temperature on the left.

\section{\label{sec:conclusions}Conclusions}

In conclusion, we have developed a generalized quantum optical noise formalism QFED that
can unambiguously describe the quantum aspects of propagating optical fields
in arbitrary stratified media, while being fully compliant with the 
canonical commutation relations.
In particular, the QFED allows calculating position-dependent photon-number
expectation values for the left
and right propagating fields and fully eliminates the anomalies of the
ladder operators in optical cavities.
In our model, the commutation relations are
therefore canonical. This implies that, in contrast to previous models that
involve anomalous commutation relations, our results do not predict, e.g.,
any observable threshold for the second harmonic generation inside
cavities as the threshold is directly linked to the ladder operators.
Experimental measurements of the existence of a second harmonic generation
threshold may therefore allow demonstrating the importance of correct
normalization of the commutation relations as well as the affiliated
normal modes. In addition, the QFED framework enables,
e.g., the separation of the quantum optical Poynting vector
and related field quantities to their left and right propagating
components using a photon-number-based presentation and the
interference density of states. In practical modeling tasks, the QFED
provides simple tools for studying optical energy transfer and the formation of thermal balance
in complex interfering nanostructures, highlighting the fundamentally nonlocal
nature of the energy transfer.

\begin{acknowledgments}
This work has in part been funded by the Academy of Finland and the Aalto Energy Efficiency Research Programme.
\end{acknowledgments}


\begin{thebibliography}{19}%
\makeatletter
\providecommand \@ifxundefined [1]{%
 \@ifx{#1\undefined}
}%
\providecommand \@ifnum [1]{%
 \ifnum #1\expandafter \@firstoftwo
 \else \expandafter \@secondoftwo
 \fi
}%
\providecommand \@ifx [1]{%
 \ifx #1\expandafter \@firstoftwo
 \else \expandafter \@secondoftwo
 \fi
}%
\providecommand \natexlab [1]{#1}%
\providecommand \enquote  [1]{``#1''}%
\providecommand \bibnamefont  [1]{#1}%
\providecommand \bibfnamefont [1]{#1}%
\providecommand \citenamefont [1]{#1}%
\providecommand \href@noop [0]{\@secondoftwo}%
\providecommand \href [0]{\begingroup \@sanitize@url \@href}%
\providecommand \@href[1]{\@@startlink{#1}\@@href}%
\providecommand \@@href[1]{\endgroup#1\@@endlink}%
\providecommand \@sanitize@url [0]{\catcode `\\12\catcode `\$12\catcode
  `\&12\catcode `\#12\catcode `\^12\catcode `\_12\catcode `\%12\relax}%
\providecommand \@@startlink[1]{}%
\providecommand \@@endlink[0]{}%
\providecommand \url  [0]{\begingroup\@sanitize@url \@url }%
\providecommand \@url [1]{\endgroup\@href {#1}{\urlprefix }}%
\providecommand \urlprefix  [0]{URL }%
\providecommand \Eprint [0]{\href }%
\providecommand \doibase [0]{http://dx.doi.org/}%
\providecommand \selectlanguage [0]{\@gobble}%
\providecommand \bibinfo  [0]{\@secondoftwo}%
\providecommand \bibfield  [0]{\@secondoftwo}%
\providecommand \translation [1]{[#1]}%
\providecommand \BibitemOpen [0]{}%
\providecommand \bibitemStop [0]{}%
\providecommand \bibitemNoStop [0]{.\EOS\space}%
\providecommand \EOS [0]{\spacefactor3000\relax}%
\providecommand \BibitemShut  [1]{\csname bibitem#1\endcsname}%
\let\auto@bib@innerbib\@empty
%</preamble>
\bibitem [{\citenamefont {Gauvin}\ \emph {et~al.}(2014)\citenamefont {Gauvin},
  \citenamefont {Collette},\ and\ \citenamefont {Beaudoin}}]{Gauvin2014}%
  \BibitemOpen
  \bibfield  {author} {\bibinfo {author} {\bibfnamefont {S.}~\bibnamefont
  {Gauvin}}, \bibinfo {author} {\bibfnamefont {M.}~\bibnamefont {Collette}}, \
  and\ \bibinfo {author} {\bibfnamefont {N.}~\bibnamefont {Beaudoin}},\ }in\
  \href {http://www.osapublishing.org/abstract.cfm?URI=LS-2014-JTu3A.29} {\emph
  {\bibinfo {booktitle} {Frontiers in Optics 2014}}}\ (\bibinfo  {publisher}
  {OSA},\ \bibinfo {year} {2014})\BibitemShut {NoStop}%
\bibitem [{\citenamefont {Collette}\ \emph {et~al.}(2013)\citenamefont
  {Collette}, \citenamefont {Beaudoin},\ and\ \citenamefont
  {Gauvin}}]{Collette2013}%
  \BibitemOpen
  \bibfield  {author} {\bibinfo {author} {\bibfnamefont {M.}~\bibnamefont
  {Collette}}, \bibinfo {author} {\bibfnamefont {N.}~\bibnamefont {Beaudoin}},
  \ and\ \bibinfo {author} {\bibfnamefont {S.}~\bibnamefont {Gauvin}},\ }\href
  {\doibase 10.1117/12.2017587} {\bibfield  {journal} {\bibinfo  {journal}
  {Proc. SPIE}\ }\textbf {\bibinfo {volume} {8772}},\ \bibinfo {pages} {87721D}
  (\bibinfo {year} {2013})}\BibitemShut {NoStop}%
\bibitem [{\citenamefont {Partanen}\ \emph
  {et~al.}(2014{\natexlab{a}})\citenamefont {Partanen}, \citenamefont
  {H\"ayrynen}, \citenamefont {Oksanen},\ and\ \citenamefont
  {Tulkki}}]{Partanen2014a}%
  \BibitemOpen
  \bibfield  {author} {\bibinfo {author} {\bibfnamefont {M.}~\bibnamefont
  {Partanen}}, \bibinfo {author} {\bibfnamefont {T.}~\bibnamefont
  {H\"ayrynen}}, \bibinfo {author} {\bibfnamefont {J.}~\bibnamefont {Oksanen}},
  \ and\ \bibinfo {author} {\bibfnamefont {J.}~\bibnamefont {Tulkki}},\ }\href
  {\doibase 10.1103/PhysRevA.89.033831} {\bibfield  {journal} {\bibinfo
  {journal} {Phys. Rev. A}\ }\textbf {\bibinfo {volume} {89}},\ \bibinfo
  {pages} {033831} (\bibinfo {year} {2014}{\natexlab{a}})}\BibitemShut
  {NoStop}%
\bibitem [{\citenamefont {Partanen}\ \emph
  {et~al.}(2014{\natexlab{b}})\citenamefont {Partanen}, \citenamefont
  {H\"ayrynen}, \citenamefont {Oksanen},\ and\ \citenamefont
  {Tulkki}}]{Partanen2014c}%
  \BibitemOpen
  \bibfield  {author} {\bibinfo {author} {\bibfnamefont {M.}~\bibnamefont
  {Partanen}}, \bibinfo {author} {\bibfnamefont {T.}~\bibnamefont
  {H\"ayrynen}}, \bibinfo {author} {\bibfnamefont {J.}~\bibnamefont {Oksanen}},
  \ and\ \bibinfo {author} {\bibfnamefont {J.}~\bibnamefont {Tulkki}},\ }\href
  {\doibase 10.1103/PhysRevA.90.063804} {\bibfield  {journal} {\bibinfo
  {journal} {Phys. Rev. A}\ }\textbf {\bibinfo {volume} {90}},\ \bibinfo
  {pages} {063804} (\bibinfo {year} {2014}{\natexlab{b}})}\BibitemShut
  {NoStop}%
\bibitem [{\citenamefont {Ueda}\ and\ \citenamefont {Imoto}(1994)}]{Ueda1994}%
  \BibitemOpen
  \bibfield  {author} {\bibinfo {author} {\bibfnamefont {M.}~\bibnamefont
  {Ueda}}\ and\ \bibinfo {author} {\bibfnamefont {N.}~\bibnamefont {Imoto}},\
  }\href {\doibase 10.1103/PhysRevA.50.89} {\bibfield  {journal} {\bibinfo
  {journal} {Phys. Rev. A}\ }\textbf {\bibinfo {volume} {50}},\ \bibinfo
  {pages} {89} (\bibinfo {year} {1994})}\BibitemShut {NoStop}%
\bibitem [{\citenamefont {Raymer}\ and\ \citenamefont
  {McKinstrie}(2013)}]{Raymer2013}%
  \BibitemOpen
  \bibfield  {author} {\bibinfo {author} {\bibfnamefont {M.~G.}\ \bibnamefont
  {Raymer}}\ and\ \bibinfo {author} {\bibfnamefont {C.~J.}\ \bibnamefont
  {McKinstrie}},\ }\href {\doibase 10.1103/PhysRevA.88.043819} {\bibfield
  {journal} {\bibinfo  {journal} {Phys. Rev. A}\ }\textbf {\bibinfo {volume}
  {88}},\ \bibinfo {pages} {043819} (\bibinfo {year} {2013})}\BibitemShut
  {NoStop}%
\bibitem [{\citenamefont {Barnett}\ \emph {et~al.}(1996)\citenamefont
  {Barnett}, \citenamefont {Gilson}, \citenamefont {Huttner},\ and\
  \citenamefont {Imoto}}]{Barnett1996}%
  \BibitemOpen
  \bibfield  {author} {\bibinfo {author} {\bibfnamefont {S.~M.}\ \bibnamefont
  {Barnett}}, \bibinfo {author} {\bibfnamefont {C.~R.}\ \bibnamefont {Gilson}},
  \bibinfo {author} {\bibfnamefont {B.}~\bibnamefont {Huttner}}, \ and\
  \bibinfo {author} {\bibfnamefont {N.}~\bibnamefont {Imoto}},\ }\href
  {\doibase 10.1103/PhysRevLett.77.1739} {\bibfield  {journal} {\bibinfo
  {journal} {Phys. Rev. Lett.}\ }\textbf {\bibinfo {volume} {77}},\ \bibinfo
  {pages} {1739} (\bibinfo {year} {1996})}\BibitemShut {NoStop}%
\bibitem [{\citenamefont {Aiello}(2000)}]{Aiello2000}%
  \BibitemOpen
  \bibfield  {author} {\bibinfo {author} {\bibfnamefont {A.}~\bibnamefont
  {Aiello}},\ }\href {\doibase 10.1103/PhysRevA.62.063813} {\bibfield
  {journal} {\bibinfo  {journal} {Phys. Rev. A}\ }\textbf {\bibinfo {volume}
  {62}},\ \bibinfo {pages} {063813} (\bibinfo {year} {2000})}\BibitemShut
  {NoStop}%
\bibitem [{\citenamefont {Di~Stefano}\ \emph {et~al.}(2000)\citenamefont
  {Di~Stefano}, \citenamefont {Savasta},\ and\ \citenamefont
  {Girlanda}}]{Stefano2000}%
  \BibitemOpen
  \bibfield  {author} {\bibinfo {author} {\bibfnamefont {O.}~\bibnamefont
  {Di~Stefano}}, \bibinfo {author} {\bibfnamefont {S.}~\bibnamefont {Savasta}},
  \ and\ \bibinfo {author} {\bibfnamefont {R.}~\bibnamefont {Girlanda}},\
  }\href {\doibase 10.1103/PhysRevA.61.023803} {\bibfield  {journal} {\bibinfo
  {journal} {Phys. Rev. A}\ }\textbf {\bibinfo {volume} {61}},\ \bibinfo
  {pages} {023803} (\bibinfo {year} {2000})}\BibitemShut {NoStop}%
\bibitem [{\citenamefont {Partanen}\ \emph
  {et~al.}(2014{\natexlab{c}})\citenamefont {Partanen}, \citenamefont
  {H{\"a}yrynen}, \citenamefont {Oksanen},\ and\ \citenamefont
  {Tulkki}}]{Partanen2014b}%
  \BibitemOpen
  \bibfield  {author} {\bibinfo {author} {\bibfnamefont {M.}~\bibnamefont
  {Partanen}}, \bibinfo {author} {\bibfnamefont {T.}~\bibnamefont
  {H{\"a}yrynen}}, \bibinfo {author} {\bibfnamefont {J.}~\bibnamefont
  {Oksanen}}, \ and\ \bibinfo {author} {\bibfnamefont {J.}~\bibnamefont
  {Tulkki}},\ }in\ \href@noop {} {\emph {\bibinfo {booktitle} {Proc. SPIE 9136,
  Nonlinear Optics and Its Applications VIII; and Quantum Optics III}}},\
  \bibinfo {series and number} {\bibinfo {number} {91362B}}\ (\bibinfo
  {publisher} {SPIE},\ \bibinfo {year} {2014})\BibitemShut {NoStop}%
\bibitem [{\citenamefont {Kn\"oll}\ \emph {et~al.}(1987)\citenamefont
  {Kn\"oll}, \citenamefont {Vogel},\ and\ \citenamefont {Welsch}}]{Knoll1987}%
  \BibitemOpen
  \bibfield  {author} {\bibinfo {author} {\bibfnamefont {L.}~\bibnamefont
  {Kn\"oll}}, \bibinfo {author} {\bibfnamefont {W.}~\bibnamefont {Vogel}}, \
  and\ \bibinfo {author} {\bibfnamefont {D.~G.}\ \bibnamefont {Welsch}},\
  }\href {\doibase 10.1103/PhysRevA.36.3803} {\bibfield  {journal} {\bibinfo
  {journal} {Phys. Rev. A}\ }\textbf {\bibinfo {volume} {36}},\ \bibinfo
  {pages} {3803} (\bibinfo {year} {1987})}\BibitemShut {NoStop}%
\bibitem [{\citenamefont {Kn\"oll}\ \emph {et~al.}(1991)\citenamefont
  {Kn\"oll}, \citenamefont {Vogel},\ and\ \citenamefont {Welsch}}]{Knoll1991}%
  \BibitemOpen
  \bibfield  {author} {\bibinfo {author} {\bibfnamefont {L.}~\bibnamefont
  {Kn\"oll}}, \bibinfo {author} {\bibfnamefont {W.}~\bibnamefont {Vogel}}, \
  and\ \bibinfo {author} {\bibfnamefont {D.-G.}\ \bibnamefont {Welsch}},\
  }\href {\doibase 10.1103/PhysRevA.43.543} {\bibfield  {journal} {\bibinfo
  {journal} {Phys. Rev. A}\ }\textbf {\bibinfo {volume} {43}},\ \bibinfo
  {pages} {543} (\bibinfo {year} {1991})}\BibitemShut {NoStop}%
\bibitem [{\citenamefont {Allen}\ and\ \citenamefont
  {Stenholm}(1992)}]{Allen1992}%
  \BibitemOpen
  \bibfield  {author} {\bibinfo {author} {\bibfnamefont {L.}~\bibnamefont
  {Allen}}\ and\ \bibinfo {author} {\bibfnamefont {S.}~\bibnamefont
  {Stenholm}},\ }\href@noop {} {\bibfield  {journal} {\bibinfo  {journal} {Opt.
  Commun.}\ }\textbf {\bibinfo {volume} {93}},\ \bibinfo {pages} {253}
  (\bibinfo {year} {1992})}\BibitemShut {NoStop}%
\bibitem [{\citenamefont {Huttner}\ and\ \citenamefont
  {Barnett}(1992)}]{Huttner1992}%
  \BibitemOpen
  \bibfield  {author} {\bibinfo {author} {\bibfnamefont {B.}~\bibnamefont
  {Huttner}}\ and\ \bibinfo {author} {\bibfnamefont {S.~M.}\ \bibnamefont
  {Barnett}},\ }\href {\doibase 10.1103/PhysRevA.46.4306} {\bibfield  {journal}
  {\bibinfo  {journal} {Phys. Rev. A}\ }\textbf {\bibinfo {volume} {46}},\
  \bibinfo {pages} {4306} (\bibinfo {year} {1992})}\BibitemShut {NoStop}%
\bibitem [{\citenamefont {Barnett}\ \emph {et~al.}(1995)\citenamefont
  {Barnett}, \citenamefont {Matloob},\ and\ \citenamefont
  {Loudon}}]{Barnett1995}%
  \BibitemOpen
  \bibfield  {author} {\bibinfo {author} {\bibfnamefont {S.~M.}\ \bibnamefont
  {Barnett}}, \bibinfo {author} {\bibfnamefont {R.}~\bibnamefont {Matloob}}, \
  and\ \bibinfo {author} {\bibfnamefont {R.}~\bibnamefont {Loudon}},\
  }\href@noop {} {\bibfield  {journal} {\bibinfo  {journal} {J. Mod. Opt.}\
  }\textbf {\bibinfo {volume} {42}},\ \bibinfo {pages} {1165} (\bibinfo {year}
  {1995})}\BibitemShut {NoStop}%
\bibitem [{\citenamefont {Matloob}\ \emph {et~al.}(1995)\citenamefont
  {Matloob}, \citenamefont {Loudon}, \citenamefont {Barnett},\ and\
  \citenamefont {Jeffers}}]{Matloob1995}%
  \BibitemOpen
  \bibfield  {author} {\bibinfo {author} {\bibfnamefont {R.}~\bibnamefont
  {Matloob}}, \bibinfo {author} {\bibfnamefont {R.}~\bibnamefont {Loudon}},
  \bibinfo {author} {\bibfnamefont {S.~M.}\ \bibnamefont {Barnett}}, \ and\
  \bibinfo {author} {\bibfnamefont {J.}~\bibnamefont {Jeffers}},\ }\href
  {\doibase 10.1103/PhysRevA.52.4823} {\bibfield  {journal} {\bibinfo
  {journal} {Phys. Rev. A}\ }\textbf {\bibinfo {volume} {52}},\ \bibinfo
  {pages} {4823} (\bibinfo {year} {1995})}\BibitemShut {NoStop}%
\bibitem [{\citenamefont {Matloob}\ and\ \citenamefont
  {Loudon}(1996)}]{Matloob1996}%
  \BibitemOpen
  \bibfield  {author} {\bibinfo {author} {\bibfnamefont {R.}~\bibnamefont
  {Matloob}}\ and\ \bibinfo {author} {\bibfnamefont {R.}~\bibnamefont
  {Loudon}},\ }\href {\doibase 10.1103/PhysRevA.53.4567} {\bibfield  {journal}
  {\bibinfo  {journal} {Phys. Rev. A}\ }\textbf {\bibinfo {volume} {53}},\
  \bibinfo {pages} {4567} (\bibinfo {year} {1996})}\BibitemShut {NoStop}%
\bibitem [{\citenamefont {Janowicz}\ \emph {et~al.}(2003)\citenamefont
  {Janowicz}, \citenamefont {Reddig},\ and\ \citenamefont
  {Holthaus}}]{Janowicz2003}%
  \BibitemOpen
  \bibfield  {author} {\bibinfo {author} {\bibfnamefont {M.}~\bibnamefont
  {Janowicz}}, \bibinfo {author} {\bibfnamefont {D.}~\bibnamefont {Reddig}}, \
  and\ \bibinfo {author} {\bibfnamefont {M.}~\bibnamefont {Holthaus}},\ }\href
  {\doibase 10.1103/PhysRevA.68.043823} {\bibfield  {journal} {\bibinfo
  {journal} {Phys. Rev. A}\ }\textbf {\bibinfo {volume} {68}},\ \bibinfo
  {pages} {043823} (\bibinfo {year} {2003})}\BibitemShut {NoStop}%
\bibitem [{\citenamefont {Loudon}(2000)}]{Loudon2000}%
  \BibitemOpen
  \bibfield  {author} {\bibinfo {author} {\bibfnamefont {R.}~\bibnamefont
  {Loudon}},\ }\href@noop {} {\emph {\bibinfo {title} {The quantum theory of
  light}}}\ (\bibinfo  {publisher} {Oxford University Press, Oxford},\ \bibinfo
  {year} {2000})\BibitemShut {NoStop}%
\end{thebibliography}
\end{document}